# Confidential Computing for Cloud Security: Exploring Hardware based Encryption Using Trusted Execution Environments

Dhruv Deepak Agarwal            Aswani Kumar Cherukuri*

School of Computer Science Engineering and Information Systems

Vellore Institute of Technology, Vellore, 632014, India.

Email: cherukuri@acm.org

**ABSTRACT:** The growth of cloud computing has revolutionized data processing and storage capacities to another levels of scalability and flexibility. But in the process, it has created a huge challenge of security, especially in terms of safeguarding sensitive data. Classical security practices, including encryption at rest and during transit, fail to protect data in use and expose it to various possible breaches. In response to this problem , Confidential Computing has been a tool ,seeking to secure data in processing by usage of hardware-based Trusted Execution Environments (TEEs). TEEs, including Intel's Software Guard Extensions (SGX) and ARM's TrustZone, offers protected contexts within the processor, where data is kept confidential ,intact and secure , even with malicious software or compromised operating systems.  In this research, we have explored the architecture and security features of TEEs like Intel SGX and ARM TrustZone, and their effectiveness in improving cloud data security. From a thorough literature survey ,we have analyzed the deployment strategies, performance indicators, and practical uses of these TEEs for the same purpose. In addition, we have discussed the issues regarding deployment, possible weaknesses, scalability issues, and integration issues. Our results focuses on the central position of TEEs in strengthening and advancing cloud security infrastructures, pointing towards their ability to create a secure foundation for Confidential Computing.

**1.INTRODUCTION:** The fast growth of cloud computing has considerably changed how businesses and individuals manages data,providing scalable, on-demand resources for everything from enterprise applications to individual user cloud storage. As cloud infrastructures and services have become the base of modern computing, it has become important to ensure data security in these environment. Traditional security measures such as encryption at rest and encryption in transit provide security and protection for data stored on physical devices or during transmission across networks but these methods are ineffective and do not work  when it comes to securing data in use, particularly in case of  cloud computing, where data is processed in memory and often exposed without any encryption.

In multitenant cloud environment, where computational resources are shared among various users, the vulnerability of attack on data in use is increased much more. For such environment data from different different users might exist on the some same physical device/ infrastructure, making it easily for an attacker  to unauthorizely access and perform attacks during data processing. This also comes up with  significant security risks, including data breaches, unauthorized access to sensitive information, and might also cause huge financial damage to both users and organizations. Therefore, protecting data during its processing phase has emerged as one of the most critical challenges in cloud security today.

In order to cope up with this concerns, the concept of Confidential Computing is been used and is been gaining attention.It aims to not only to protect data at rest and in transit but also to protect data during processing, for this purpose it make use of Trusted Execution Environments (TEEs). A TEE is an isolated, secured area which exists within a processor and it ensures that sensitive data and code remain protected from unwanted unauthorized access. TEEs such as Intel Software Guard Extensions (SGX) and ARM TrustZone offers a very good and effective solution to safeguard sensitive data in cloud,ensuring that even if the larger system is compromise, the data remains protected[9],[13],[16].

Integration of TEEs into cloud services by top providers like Amazon Web Services, Microsoft Azure, and Google Cloud is a major step towards trusted cloud model deployment. Presently, these cloud providers provide containers with TEE support to ensure that sensitive workload can be securely executed. Studies have shown the effectiveness of TEEs in isolating workloads, preventing unauthorized access, and secure execution of critical operations. Intel SGX has shown effectiveness in protecting encrypted machine learning models, while ARM TrustZone is used in mobile and edge computing to secure data even it is in use.

The use of TEEs is not limited to cloud computing alone. Other Areas such as artificial intelligence (AI), big data analytics, and Internet of Things (IoT) are also seeing usage of Confidential Computing. In AI TEEs can be used to safeguard sensitive datasets and model parameters while training and during deployment. AI workloads tend to deal with highly sensitive data eg medical records, financial information, and personal identifiers etc. which must be protected while using. Recent studies, such as [7], investigate the practical implications of securing AI inference workloads using a hybrid CPU-GPU setup in a confidential computing environment. The findings suggest that while Intel TDX can safeguard the CPU-side of computation, the absence of fully confidential GPU execution paths presents a challenge, particularly when deploying large models like BERT in cloud environments.

For field like big data analytics, TEEs are being utilized to secure workflows and analytical queries, as explored in studies like DuckDB-SGX2[18], which addresses confidential analytical query processing. Similarly, research on workflow scheduling algorithms, such as SGX-E2C2D[2], shows how TEEs can ensure data confidentiality throughout distributed computing pipelines in the cloud[7],[15].

Confidential Computing is also increasing its impact into Internet Of Things(IoT) ecosystems, which generate a large amount of sensitive data. And in scenarios such as smart healthcare, industrial automation, and connected vehicles , to secure data end-to-end from device collection to cloud processing is very much important. Recent work such as [1], shows how TEEs can be integrated into cloud-assisted IoT architectures for robust data protection throughout the data lifecycle.

To simplify the programming and implementation of TEEs, frameworks such as LuaGuardia[4] and HasTEE/HasTEE+ [5],[6] have being developed. This platforms allow developers to build secure cloud-native applications without needing great low-level security expertise. At the same time, the industry usage of TEE framework is being explored to create standardized toolkits for deployment of secure cloud applications across various TEE-capable hardware platforms.

Even though there are many advantages, Trusted Execution Environments (TEEs) also comes up with some drawbacks. They can have probability of having performance overheads, limited memory,complex and difficult debugging, and platform compatibility issues. Studies such as TS-Perf provide critical benchmarks and analysis to understand the limitations and trade-offs involved in using TEEs [9]. Furthermore, researchers are actively working to ensure code confidentiality and reduce attack surfaces in TEE applications, particularly as threats like side-channel attacks continue to become more complex [10], [11].

Rather than technical limitations, Trusted Execution Environments (TEEs) also suffers with architectural and system challenges. One of the major concern is the lack of standardized, cross-platform development frameworks, making it difficult to make use of portable TEE applications across different different cloud vendors and hardware architecture [4], [6], [20]. TEEs also struggle with handling input/output securely , while data inside may be protected, data entering or leaving has the chances of being exposed to leakage or can be hampered unless it is specially secured[10],[11]. Furthermore, attestation mechanisms, which prove the authenticity and integrity of enclaves, remain fragmented and non-uniform across platforms, limiting interoperability and trust between users [19], [20]. Researches also indicates that TEE-based applications are open to rollback and replay attacks,especially in distributed systems where synchronization secure state across regions is more complex [9], [10]. Finally, using TEEs in large cloud and edge systems isn't easy ,it needs proper coordination, because managing the life cycle of different enclaves, setting them up, and handling secure keys puts extra work on both developers and system administrators [2],[9].

TEEs provide several business advantages:

- Improved Security: Protect and Secure data during computation.

- Regulatory compliance is a key driver for Confidential Computing, as it enables alignment with data protection laws such as the General Data Protection Regulation (GDPR) and the Health Insurance Portability and Accountability Act (HIPAA) [1], [14], [20]."

- Increased Trust: Enable remote attestation to validate the integrity and authenticity of processing environments.

- Innovation : Allow enterprises to handle sensitive datasets in the cloud securely without any compromise with security.

Growing applications such as decentralized AI, multi-party computation, and edge computing highlights the increasing importance and need of Confidential Computing. As organizations migrates and are shifting towards multi-cloud and hybrid architectures, the demand for hardware-based Trusted Execution Environments (TEEs) is expected to increase more and more. This advancements in computing makes it necessary to make use of TEEs to ensure the confidentiality, integrity, and privacy of data across distributed and different mixed kind of environments [1], [3], [12], [13].

## 2. LITERATURE REVIEW:

| Paper Title | Focus Area | Security Mechanism | Target Platform | Performance Metrics | Evaluation Method |
| --- | --- | --- | --- | --- | --- |
| Confidential Computing Architectures for Enhanced Data Security in Cloud Environments | Secure cloud architecture for big data analytics and AI using confidential computing to protect data, logic, and computation in transit, at rest, and in use | Encrypted containers (Singularity) Encrypted virtual machines (VMs) using AMD SEV (Secure Encrypted Virtualization) Encrypted storage (Quobyte with AES-XTS) Resource isolation (OpenStack) | Cloud environments leveraging OpenStack for resource provisioning AMD SEV-enabled hardware for VM encryption Singularity containers for workflow encapsulation Quobyte storage systems for encrypted data at rest | I/O bandwidth (measured using IOR benchmark) Execution time (measured using BPMF benchmark inside Singularity containers) Impact of encryption on storage and container performance | Benchmarking I/O bandwidth with and without storage encryption (IOR) Benchmarking execution time with and without container encryption (BPMF) Preliminary analysis of performance overhead introduced by encryption mechanisms |
| Evaluation of confidential computing for AI inference in cloud environments using CPU-GPU Trusted Execution Environments (TEEs), specifically assessing performance and security trade-offs for large language models (LLMs) on modern hardware | Evaluation of confidential computing for AI inference in cloud environments using CPU-GPU Trusted Execution Environments (TEEs), specifically assessing performance and security trade-offs for large language models (LLMs) on modern hardware | Trusted Execution Environments (TEEs) for both CPU (Intel TDX) and GPU (NVIDIA H100 Hopper architecture) to provide hardware-based data confidentiality and integrity during AI inference. Attestation mechanisms for verifying the integrity and authenticity of the TEE and the workloads running within | Intel TDX-enabled CPUs. NVIDIA H100 GPUs with confidential computing capabilities | Overhead introduced by confidential computing in CPU-GPU TEEs compared to non-confidential setups[1]. Sensitivity of LLM inference performance to model type and batch size[1]. Throughput and latency for real-world AI workloads (e.g., BERT, LLaMA, Granite models) | Comparative analysis of confidential vs. non-confidential setups Assessment of attestation and key management integration in cloud environments |

| | | it | | | |
|---|---|---|---|---|---|
| Towards Confidential Computing: A Secure Cloud Architecture for Big Data Analytics and AI | Secure cloud architecture for big data analytics and AI using confidential computing | Hardware-based Trusted Execution Environments (TEEs)<br><br>Data encryption in use (memory and processing) | Cloud environments (public, private, hybrid)<br><br>Edge devices<br><br>Hardware platforms such as Intel TDX, AMD SEV, Arm CCA, GPUs, FPGAs | Security level (confidentiality, integrity)<br><br>Computational overhead<br><br>Scalability | Threat modeling and attack analysis<br><br>Security testing of TEEs<br><br>Performance benchmarking (overhead, scalability) |
| Ensuring End-to-End IoT Data Security and Privacy Through Cloud-Enhanced Confidential Computing | End-to-end IoT data security and privacy using cloud-enhanced confidential computing, specifically analyzing and defending against side-channel attacks on Trusted Execution Environment (TEE)-based IoT solutions | Trusted Execution Environments (TEEs)<br><br>Data oblivious execution<br><br>Padding techniques to obscure memory access patterns | IoT devices transmitting data to untrusted cloud services, with security mechanisms deployed in the cloud using TEEs | Accuracy of device type detection (attack success rate)<br><br>Accuracy of value prediction (attack success rate) | Demonstration of side-channel attacks exploiting memory access patterns on TEE-based IoT solutions<br><br>Empirical measurement of attack accuracy before and after applying proposed defenses |
| Enabling Rack-scale Confidential Computing using Heterogeneous Trusted Execution Environment | Design and implementation of a rack-scale Trusted Execution Environment (TEE) for scalable | Heterogeneous Trusted Execution Environment (HETEE) combining | Rack-scale heterogeneous computing infrastructure in cloud data | Latency overhead for inference: approximately 14.24%. | Implementation of the HETEE prototype system. |
| | confidential computing in heterogeneous cloud/data center environments.<br><br>Supporting large-scale confidential AI workloads, such as deep neural network (DNN) training and inference, with minimal performance overhead. | hardware TEEs like Intel SGX, AMD SEV, ARM TrustZone with a centralized Security Controller (SC) hardware module.<br><br>Use of enclaves for secure isolated execution on CPUs and Data Streaming Accelerators (DSAs). | centers.<br><br>Includes CPU-TEE nodes, DSA-TEE nodes, and non-TEE legacy nodes connected via PCIe within a physically hardened rack container. | Throughput overhead for training: approximately 0.60%. | Performance evaluation using large-scale DNN training and inference benchmarks. |
| Privacy-Preserving Decentralized AI with Confidential Computing | Privacy-Preserving Decentralized AI with Confidential Computing | Confidential computing (data remains encrypted during computation)<br><br>Secure multi-party computation<br><br>Differential privacy | Decentralized AI systems and federated learning environments.<br><br>Distributed devices or organizations participating in collaborative AI | Scalability<br><br>Resilience<br><br>Efficiency of federated learning protocols | Theoretical analysis of privacy guarantees<br><br>Protocol and system architecture proposals<br><br>Analysis of secure aggregation and computation integrity |

| Title | Description | Security Mechanisms | Technical Environment | Performance/Trade-offs | Evaluation |
|---|---|---|---|---|---|
| HyperEnclave: An Open and Cross-platform Trusted Execution Environment | Design and implementation of HyperEnclave, an open and cross-platform process-based Trusted Execution Environment (TEE) that uses widely available virtualization extensions and TPM for isolation and root of trust | Isolation via virtualization extensions (e.g., VMX root and non-root modes)  Trusted software layer called RustMonitor managing enclave isolation and memory (page tables and page faults) | Isolation via virtualization extensions (e.g., VMX root and non-root modes)  Trusted software layer called RustMonitor managing enclave isolation and memory (page tables and page faults) | Overhead measured on real hardware and applications, showing small performance overhead compared to native execution  Flexible enclave operation modes to optimize security and performance trade-offs for different workloads (compute-intensive, I/O-intensive, memory-intensive) | Implementation on commodity hardware with real application benchmarks  Security evaluation through formal verification of critical components like RustMonitor and page table management |
| Secure Distributed Computing in Cloud Using Trusted Execution Environments | Secure distributed computing in cloud environments using Trusted Execution Environments (TEEs) | Hardware-based isolation (secure eclaves/TEEs)  Cryptographic mechanisms for confidentiality and integrity  Attestation for verifying enclave state and authenticity | Cloud infrastructure (public, private, hybrid)  Distributed applications running on cloud or on-premises servers  Specific implementations like Intel SGX, Intel TDX | Overhead introduced by TEE operations (e.g., encryption, attestation)  Confidentiality and integrity guarantees  Scalability and usability in distributed settings | Formal security analysis (e.g., noninterference, integrity properties)  Experimental evaluation using real TEE implementations (e.g., Intel TDX, SGX) |
| TS-Perf: General Performance Measurement of Trusted Execution Environment and Rich Execution Environment on Intel SGX, Arm TrustZone, and RISC-V Keystone | General performance measurement of Trusted Execution Environment (TEE) and Rich Execution Environment (REE) on Intel SGX, Arm TrustZone, and RISC-V Keystone. | Trusted Execution Environments (TEE) such as Intel SGX enclaves, Arm TrustZone secure world, and RISC-V Keystone secure enclaves. | Intel SGX (Software Guard Extensions)<br><br>Arm TrustZone<br><br>RISC-V Keystone | Execution time overhead<br><br>Memory usage and footprint<br><br>Context switch latency between TEE and REE | Benchmarking with standardized workloads<br><br>Micro-benchmarks for specific TEE operations (e.g., enclave entry/exit)<br><br>Comparative performance analysis across platforms |
| Confidential Computing and Related Technologies: A Critical Review | Confidential computing technologies for securing data in use<br><br>Hardware-based Trusted Execution Environments (TEEs)<br><br>Security for cloud, hybrid, and edge computing environments | Trusted Execution Environments (TEEs) / Secure<br><br>Hardware-level encryption (runtime/memory encryption)<br><br>Attestation mechanisms for environment and application trust | Public cloud, hybrid cloud, and edge computing<br><br>Virtual machines, containers, and bare-metal servers<br><br>Supported by major cloud providers (AWS Nitro, Google Cloud, Azure, Intel SGX, AMD SEV) | Data confidentiality and integrity<br><br>Attestation success/failure rates<br><br>Overhead/latency introduced by enclave/isolation | Security analysis of TEE/enclave isolation and attestation<br><br>Performance benchmarking (latency, throughput)<br><br>Compliance verification with data privacy regulations |

By reviewing papers i founded that all of them showed different ways to improve cloud security using confidential computing. These include using encrypted containers, virtual machines with AMD SEV, and secured AI models using TEEs. The main aim for all of them was to keep data safe while it is being processed and when data is in use. Many studies also tested how this different ways affected performance, speed, and how well they perform at large scale. Emerging Technologies like Intel SGX, AMD SEV, HyperEnclave, and HETEE showed slowing down of things , but they still provide a strong protection by keeping data isolated and safe. And finally, the use of special tools like attestation and memory protection in public and hybrid clouds showed that confidential computing is becoming more advanced and useful in real-world cloud systems day by day.

# 3. Analysis and Discussion

Confidential computing has come up as a very important and very beneficial improvement for security of data in cloud environments. By reviewing and gathering information from the literature review, this section present comparison and analysis of three major Trusted Execution Environments (TEEs) and also focuses on their integration within cloud infrastructure, also investigates hardware-level encryption mechanisms, and evaluates their practical implications on performance, security, and deployment.

## 3.1 Comparative Analysis of Trusted Execution Environments

The three most widely adopted TEEs in confidential computing—Intel SGX, ARM TrustZone, and RISC-V Keystone—offer distinct architectures and security guarantees. Table I summarizes the key differences between these platforms, based on the reviewed literature.

Table I: Comparative Overview of TEE Architectures

| Feature | Intel SGX | ARM TrustZone | RISC-V Keystone |
|---|---|---|---|
| Isolation Scope | Application level enclaves: Secured regions for trusted execution, isolated from the host Operating System and other applications.[9],[18] | Entire secure world: Isolates critical operation and data from the normal world using ARM's secure world[5],[13]. | Flexible user-space enclaves: Provides customization isolation for specific applications in a modular manner[16],[20]. |
| Memory Protection | EPC with encryption and integrity[9], [18] | Hardware firewalls: Controls and isolates memory regions between secure and normal worlds to protect sensitive data.[5],[13] | PMP-based isolation (Physical Memory Protection): Ensures memory regions for trusted applications are secured with configurable protection.[16],[20] |
| Performance Overhead | Moderate to high (varies with workload)[9], [18] | Low: Minimal overhead due to simple isolation between secure and normal worlds.[5],[9] | Configurable: Overhead can be adjusted based on specific workload requirements, with flexibility for optimized performance.[16],[20] |
| Availability | Available on mainstream Intel CPUs: Supported by a wide range of Intel processors, particularly in cloud environments.[9], [18] | Embedded and mobile SoCs: Used in mobile devices, embedded systems, and IoT applications.[5],[13] | Research and open-source hardware: Still in early research phases, with support for open-source RISC-V hardware platforms.[16],[20] |
| Use Case Focus | Cloud applications, AI inference: Particularly beneficial for cloud environments and processing sensitive data in AI models.[9], [18] | IoT, mobile security: Targeted for securing IoT devices and mobile applications where minimal overhead is critical[5],[13]. | Research, modular security design: Focused on providing flexibility and modularity for research and experimental applications.[16],[20] |
| Security Attacks Reported | Side-channel attacks, rollback attacks: Vulnerable to timing-based and other side-channel attacks, and attacks on enclave integrity. [9], [18] | Interface abuse, firmware-level attacks: Potential vulnerabilities arise from interface manipulation and firmware-level attacks.[9] | Fewer attacks reported, but still under scrutiny: While fewer attacks have been reported, Keystone is still under research for potential vulnerabilities[16], [19] |

From this comparison, it becomes clear that no TEE offers a perfect solution each comes with its strengths and weakness, Intel SGX leads in commercial deployment but suffers from well-known side-channel vulnerabilities and attacks . TrustZone, though simpler, lacks fine-grained control over applications, while Keystone offers research flexibility but lacks mainstream support.

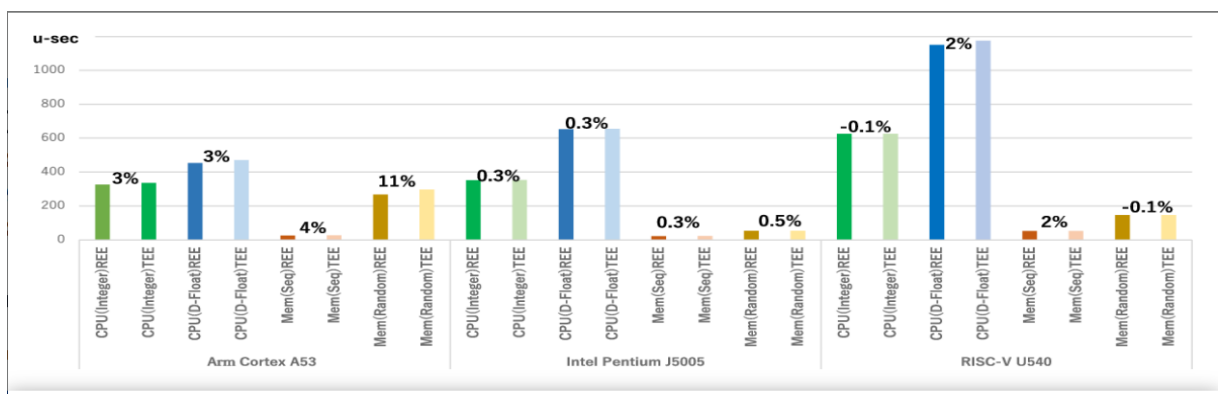

**Fig. 1. Intel SGX memory protection mechanism. Adapted from [9].**

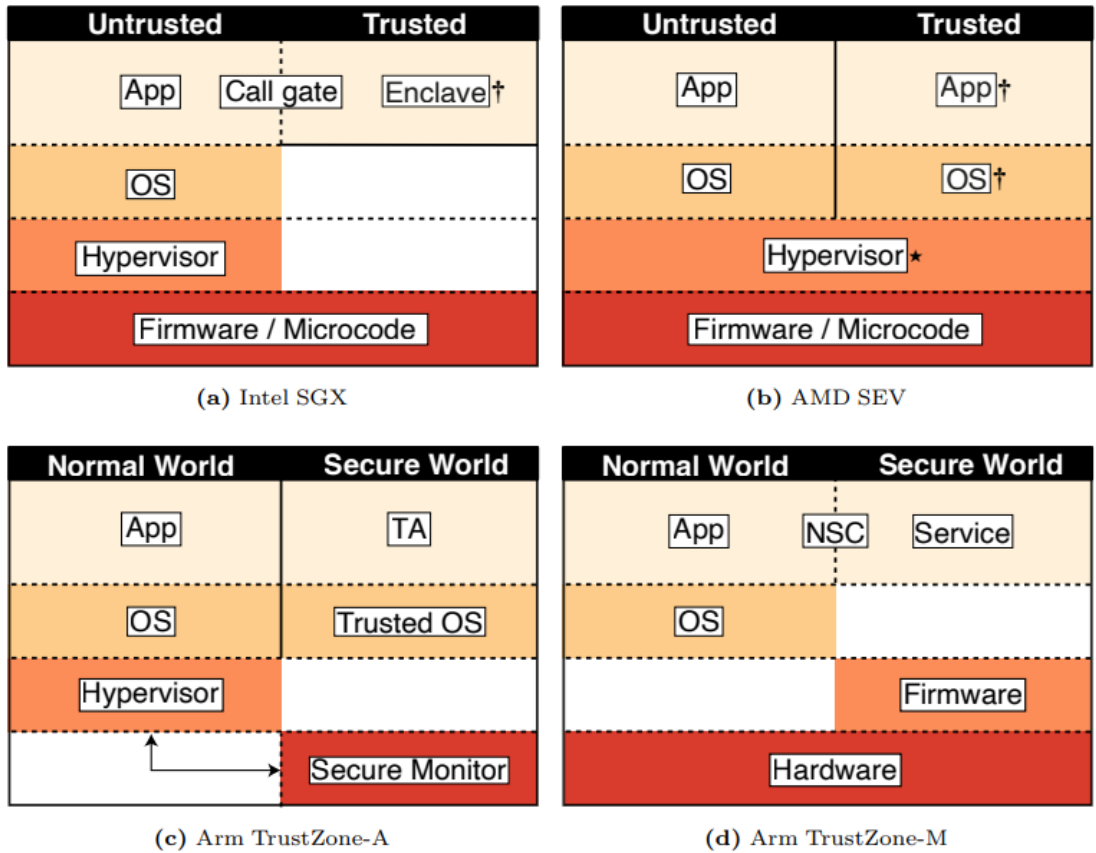

**Fig. 2.** ARM TrustZone architecture showing division of secure and normal worlds. Adapted from [19].

### 3.2 Confidential Computing in Cloud Infrastructure

Cloud provider like Microsoft Azure, Google Cloud, and AWS have introduced confidential computing instances in their services to offer hardware based isolation for sensitive computations. Azure Confidential Computing utilizes Intel SGX-enabled VMs to support secure multiparty computation and machine learning model inference [14].

Key benefits observed include:

- Runtime confidentiality: Blocks even privileged cloud administrators' access[14],[15].
- Attestation services: Enable clients to validate enclave integrity prior to provisioning data[12],[19].
- Scalable security: Supports multi-tenant workloads with isolation[13],[16].

However, challenges remain in terms of:

- Orchestration with containers (e.g., Kubernetes compatibility)[14],[16].
- Debugging and performance monitoring limitations[9],[5].
- Limited memory (128MB EPC in SGX), requiring paging[14],[15].

## 4.3 Hardware-Based Encryption Mechanisms

Confidential computing uses special hardware to keep data safe even while it's being used, not just when it's stored or is in transit. Some of the key hardware features include:

- Memory encryption engines: These are inbuilt in computer chips and make sure that anything stored in RAM is encrypted automaticaly. This helps to stop hackers who might try to steal data [14],[15].

- On-chip key storage: The chip has a secured place to keep secret encryption keys. This is called a hardware root of trust, and it makes sure that the keys are safe and can't be stolen easily and are not easily accessible to unauthorized person[1],[9].

- Attestation protocols: Before any sensitive data is decrypted, these protocols check if the program and system running the data are genuine and not tampered with. This helps make sure data is only given to secure and trusted environments[8],[15],[19].

**Table II shows an evaluation of selected encryption schemes in literature.**

| Paper | Encryption Type | Key Management | TEE Used | Performance Impact |
|---|---|---|---|---|
| DuckDB-SGX2 [18] | EPC Memory Encryption | Hardware-attested KMS | Intel SGX | 10–15% latency |
| TS-Perf [9] | In-Memory AES Encryption | TPM-Backed Keys | SGX, TrustZone | Minimal |
| HasTEE [5] | On-chip Custom AES | Software Key Vault | Keystone | Low on RISC-V |
| HyperEnclave [13] | Full-stack AES-GCM | Local Hardware KMS | Intel SGX | Moderate |

Table II: Hardware Encryption Techniques in Confidential Cloud

The above table compares different encryption techniques and their associated performance impacts in the context of various Trusted Execution Environments (TEEs). As we can observe DuckDB-SGX2 [18] utilizes EPC memory encryption with hardware attested Key Management Services (KMS) on Intel SGX, which results in a 10–15% latency overhead.On other hand, TS-Perf [9] uses in memory AES encryption with TPM backed keys for SGX and ARM TrustZone giving high performance and low impact. HasTEE [5], which works on chip custom AES encryption with a software key vault, shows very low performance overhead on RISC-V Keystone. Lastly, HyperEnclave [13] applies full-stack AES-GCM encryption with a local hardware KMS on Intel SGX, coming up with a moderate performance impact. These variations in performance shows the trade-offs between enhancing security and processing efficiency in different TEE architectures.

## 4.4 Case Studies from Literature

### 4.4.1 DuckDB-SGX2

The study about DuckDB-SGX2 [18] throw light on its key contribution in integrating DuckDB a analytical engine within an Intel SGX enclave. This method was indeed discussed in paper[18], where DuckDB-SGX2 works within an enclave, ensuring the confidentiality of both the query execution and data even from the host system.

Suggestion:As the solution provided in the paper comes up with a performance trade-offs showing an increase in latency (10-15%). This performance cost must be a crucial thing to consider when evaluating the practical feasibility of using SGX for such use cases.

### 4.4.2 TS-Perf

The study about TS-Perf [9] show us the benchmarking performance trade-offs between Intel SGX, ARM TrustZone, and RISC-V Keystone. We can see that TS-Perf's key contribution is providing a suitable solution to evaluate the performance impact of these TEEs, particularly the overhead caused by context switches and secure memory access etc.

Suggestion: As in the original paper[9], we can observe TS-Perf also highlights how each TEE handles specific workloads, and it discusses specific metrics such as latency and performance under real-world conditions we could emphasize on the benchmarking suite and the focus on real-world performance to strengthen the connection.

### 4.4.3 HyperEnclave

The study on HyperEnclave [13] emphasizes focuses on simplification of the deployment of SGX in cloud native environments, specially its integration with technologies like Docker and Kubernetes. The support for attestation protocols and its impact on secure cloud-native computing are indeed key aspects of the HyperEnclave system, which aligns with the discussion as seen in the paper.

Suggestion:Felt a need to mention how HyperEnclave's develper friendly features help streamlining confidential computing deployments and helps to address challenges such as secure enclave management and orchestration which seemed to be mising in the study. Also, a mention of the performance trade-offs due to the added complexity of integration is needed

### 4.5 Security Threats and Mitigation

Trusted Execution Environments (TEEs) are beneficial but they still remain vulnerable to several attacks:

**Attacks:**

- **Side-channel attacks:** Studies from [3],[9] shows that SGX is susceptible to cache timing and branch prediction attacks where attackers can get and access sensitive data based on timing variations .

- **Firmware manipulation in ARM TrustZone**: Research from [13] shows that ARM TrustZone can be compromised if attackers/hacker manipultes the firmware running within the secure world and hence bypassing its security.

- **Roll-back and replay attacks:** Studies from [3] shows that SGX is prone to roll-back and replay attacks , where attackers can revert systems to insecure states or replay captured secure operations.

**Mitigations(Techniques and Tools Used to Eliminate Vulnerability):**

- **Constant-time coding and randomization**: These techniques can help to mitigate side-channel attacks by ensuring uniform execution time and randomized data access patterns[11].

- **Versioned sealing in SGX**: SGX uses **versioned sealing** to detect roll-backs by ensuring encrypted data matches the expected version[9].

- **Secure boot chains in TrustZone**: **Secure boot** ensures only trusted firmware is executed, protecting against firmware manipulation attacks[5].

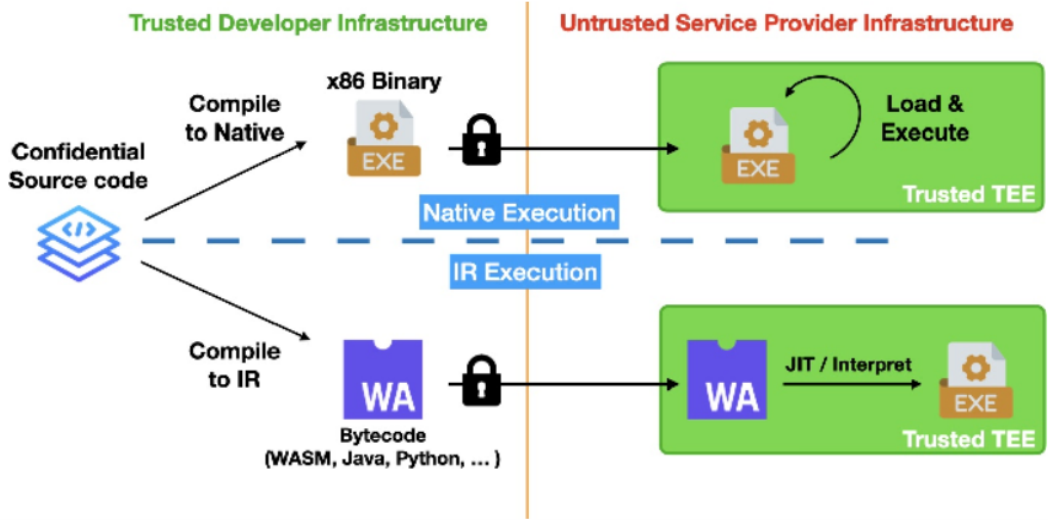

**Fig. 3. Approaches to Providing Code Confidentiality with TEEs. Adapted from [11].**

### 4.6 Summary of Observations

This section illustrates that while TEEs significantly improve cloud data confidentiality, their design choices affect scalability, security assurance, and usability and performance. Table III summarizes key comparative outcomes:

**Table III: Summary of TEE Strengths and Weaknesses**

| TEE | Strengths | Weaknesses |
|---|---|---|
| Intel SGX | Commercial-ready, Attestation support | Side-channel risks, EPC limits |
| ARM TrustZone | Broad SoC adoption, Minimal overhead | Weak isolation granularity |
| Keystone | Open-source, Flexible architecture | Not production-grade, limited adoption |

To summarize, Intel SGX is widely used in commercial applications because of it ability to support attestation which helps in maintaining the integrity of sensitive data. Taking all this into account, I believe TEEs have a promising role in cloud security, but they clearly need more standardization and practical support tools .However it also have some security issues like side-channel attacks and its memory capacity is limited as stated in various studies. ARM TrustZone can be seen to be commonly used in system-on-chip (SoC) device and it has low performance overhead making it good for devices with limited resources. However, its isolation is not as detailed, which can create a prolem in complex cases. Keystone, an open-source TEE, is flexible and customizable but is still in the research stage and not yet ready to be used for large-scale use. Personally, I think future research should focus not just on making TEEs secure, but also easier to use and debug in real deployments.

## 5. CHALLENGES AND FUTURE RESEARCH DIRECTIONS

Confidential computing is an emerging field that helps and provides strong security guarantees, but it also comes up with several technical and operational challenges as listed below:

- Limited Enclave Memory: TEEs like Intel SGX provide only about 128MB of secure memory, which creates performance bottlenecks, especially for high memory required tasks[18].

- Performance Issues:Encryption and context switching can slow down system performance , which is a huge problem for high-performance environments [9]..

- Lack of Standardized APIs: As studies suggest that there is a lack of unified APIs ,Different TEEs make it difficult to create a single programming model, making it difficult for developers to deal with them [5].

- Difficult debugging: Studies suggest that It's difficult to debug code inside TEEs, which can make development take longer and lead to more blunders[9].

**Future Research Directions:**

- Unified TEE Frameworks: There is a need to work on cross-platform SDKs for confidential computing which may reduce fragmentation and facilitate broader support.

- Hybrid TEE Models: Need to Combine features from different TEEs so as to offer enhanced customization and better performance.

- Post-Quantum Cryptography Integration: Need to work on encryption mechanisms for future dangers threats which can be caused by quantum computing which is critical for the long-term security of TEEs.

- Scalable Confidential ML: Need to find out the use of secure enclaves for training and deployment of large ML models which can ensure the privacy of data used in AI apps.

- Cloud-native TEE Orchestration: Need for Improving the compatibility of TEEs with cloud-native technologies like Kubernetes, Docker, and serverless architectures which will improve their usefulness in upcoming modern computing environments.

## 6.CONCLUSION

In conclusion, both TEEs Intel SGX and ARM TrustZone play important roles in increasing and upholding the security and privacy of cloud computing. They both provide security for the data which is in use. While Intel SGX is more suitable and effective for high-security, data-intensive applications , both of which requires strong isolation and attestation,Whereas ARM TrustZone provides a more resource-efficient solution, suitable for mobile and edge devices. Both these TEEs together makes sure that sensitive data can be processed in the cloud without any attacks securely mitigating all the risks associated with data breaches and unauthorized access which can effect both individual and an organization. In future,As the demand for secure cloud computing continues to grow rapidly, the usage and adoption of these TEEs will play an important role in shaping the future of secure cloud environments.